\documentclass[preprint,12pt]{elsarticle}
\usepackage[T1]{fontenc}

\usepackage{amssymb}

\journal{Physics Letters A}

\usepackage{hyperref}
\usepackage{braket}
\usepackage{multirow}
\usepackage{booktabs}
\usepackage[font=footnotesize]{subcaption}
\usepackage{xcolor}
\usepackage[export]{adjustbox}
\usepackage{amssymb}
\usepackage{amsmath}
\usepackage{setspace}
\usepackage{babel}
\usepackage{comment}
\usepackage{lipsum}
\usepackage{subcaption}
\usepackage{afterpage}
\usepackage{caption}
\usepackage{float}
\usepackage{booktabs}
\usepackage{natbib}
\usepackage{soul}
\usepackage{hyperref}

\begin{document}

\begin{frontmatter}



\title{Optimization of the Memory Reset Rate of a Quantum Echo-State Network for Time Sequential Tasks}


\author[add1,add3]{Riccardo Molteni}

\author[add3]{Claudio Destri}

\author[add1,add2]{Enrico Prati\corref{cor1}}
\ead{enrico.prati@cnr.it}

\address[add1]{Istituto di Fotonica e Nanotecnologie, Consiglio Nazionale delle Ricerche, Piazza Leonardo da Vinci 32, Milan, I-20133, Italy}
\address[add2]{Dipartimento di Fisica ``Aldo Pontremoli'', Universit\`{a} degli Studi di Milano, Milan, I-20133, Italy}
\address[add3]{Dipartimento di Fisica ''G. Occhialini'', Universit\'a degli Studi di Milano Bicocca, Milan, I-20133, Italy}
\cortext[cor1]{Corresponding author}

\begin{abstract}
Quantum reservoir computing is a class of quantum machine learning algorithms involving a reservoir of an echo state network based on a register of qubits, but the dependence of its memory capacity on the hyper\-parameters is still rather unclear. In order to maximize its accuracy in time--series predictive tasks, we investi\-gate the relation between the memory of the network and the reset rate of the evolution of the quantum reservoir. 
We benchmark the network performance by three non--linear maps with fading memory on IBM quantum hardware. The memory capacity of the quantum reservoir is maximized for central values of the memory reset rate in the interval $\mathbf{[0,1]}$. 
As expected, the memory capacity increases approximately linearly with the number of qubits. After optimization of the memory reset rate, the mean squared errors of the predicted outputs in the tasks may decrease by a factor $\sim 1/5$ with respect to previous implementations.
\end{abstract}




\end{frontmatter}



\section{Introduction}\label{sect:introduction}
	
	
The continuous development of quantum computers \cite{rotta2017quantum}\cite{ferraro2020all}\cite{nielsen2010quantum}\cite{preskill2018quantum}\cite{barends2014superconducting}\cite{boixo2018characterizing} has found in machine learning one of its natural applications \cite{rocutto2021quantum}\cite{maronese2021continuous}. In particular, the field of quantum machine learning (QML) has been widely explored mainly through hybrid quantum-classical algorithms which take advantage of both quantum and classical parts during the training. Many of such applications involve variational algorithms to solve classification problems \cite{lazzarin2022multi}\cite{maronese2022quantum}, data reconstruction \cite{rocutto2021quantum} and generative adversarial networks\cite{agliardi2022optimal}. Diffe\-rently, quantum reservoir computing (QRC) deals primarily with sequential temporal tasks where the fading memory of the system plays a crucial role.
Based on the concept of reservoir computing \cite{Soriano} originally embodied by echo state networks \cite{jaeger2001echo} and liquid state machines \cite{maass2004computational} respectively, quantum reservoir computing was first proposed in 2017 by Fujii and Nakajima \cite{PhysRevApplied.8.024030}. There, an encoding of a quantum reservoir on a register of qubits is proposed. The implementation consisted of the simulation of a NMR machine \cite{fujii2021quantum, nakajima2019boosting,tran2021learning}. Other proposals based on gate-model hardware have followed \cite{chen2020temporal,dasgupta2020designing}. 
In QRC, instead of usual rate-neurons, a register of qubits provides the reservoir of an echo-state network  \cite{jaeger2004harnessing,lukovsevivcius2009reservoir,jaeger2001echo,jaeger2002tutorial},  thus realizing a quantum ESN (qESN). By a suitable quantum evolution of the system, the reservoir expands non-linearly the input values in a high dimen\-sional system so that it is sufficient to add a final linear readout to make the network capable of emulating any nonlinear map. Furthermore, the reservoir retains information about past inputs in its quantum state enabling the network to tackle temporal tasks for which memory is essential.

Quantum reservoir computing, and in particular quantum echo-state net\-works, appear as promising implementations of quantum machine learning for temporal tasks, but a detailed analysis of the dependence of the  memory properties of a qESN on its hyperparameters is still largely un\-adressed. With the spirit of filling such gap, we study the memory dependence on the reset rate which regulates the probability with which the quantum reservoir resets its internal state during the evolution.  
	
Previously, some of the authors explored QML through quantum annealers for unsupervised learning \cite{rocutto2021quantum}, quantum generative adversarial net\-works \cite{agliardi2022optimal,agliardi2022quantum}, quantum  feed-forward neural networks \cite{maronese2022quantum} and quantum variatio\-nal tensor net\-works for multiclass supervised learning \cite{lazzarin2022multi}. 
Here we explore supervised learning of time-series by a qESN through an embodiment on gate model quantum computers. The implementation of the qESN used in our work is particularly suited for noisy intermediate-scale quantum (NISQ) computers. In particular we take in consideration the subclass of universal QRs which evolves with a resettable convex linear combination of complete trace preserving (CPTP) maps, introduced in Ref. \cite{chen2020temporal}.  

We investigate the memory of the reservoir by employing the memory capacity (MC) metric \cite{PhysRevApplied.8.024030}. The memory capacity quantifies the short-term memory of the reservoir by testing its ability to reproduce previously received input values by using its current internal state. In particu\-lar, in Ref. \cite{PhysRevApplied.8.024030} an empirical evaluation suggested that a qESN of 5 qubits could exhibit a higher memory capacity than a classical ESN of 500 nodes. 

In our experi\-ments we measured the memory capacity for different values of the reset rate $\epsilon$, a fundamental hyperparameter of the QR subclass studied in this work, which regulates the rate at which the quantum state of the reservoir is reset during the time evolution, thus driving its fading memory behaviour. We compare the memory capacity of different reservoirs based on 3, 5 and 7 qubits, respectively. 

Furthermore, the value of $\epsilon$ influences also the overall performance of the network. We tested our qESN over three different tasks, all of them requiring the emulation of a nonlinear map with fading memory, that is with outputs which depend increasingly less on previous inputs as time flows. We employed two, relatively simple, second-order maps and a more involved realistic map consisting in a pair of differential equa\-tions related to the simplified dynamics of an aircraft. For all of the tasks we injected into the system random input variables and trained the network to predict the exact output values for input data not used during the training, according to the specific map. For each task we tested the implementation on a simulator with different values of $\epsilon$. Next we selected the optimal one to run the experiments on quantum hardware.
	
    
	
	
The qESN is realized via quantum circuits on both the Qiskit simulator and the IBM \texttt{ibmq\_casablanca} quantum hardware.
	
The results of the memory investigation show that the MC is maximized for the central values of $\epsilon\in [0,1]$, while it sensibly decreases for $\epsilon > 0.5$. 
Furthermore for the optimal $\epsilon=0.5$ the MC increases approximately linearly with the number of qubits of the reservoir, confirming the result obtained in Ref. \cite{PhysRevApplied.8.024030} with a different evolution for the reservoir by using a simulated NMR machine.

The qESN is able to predict outputs which match those expected with low errors, measured as normalized mean squared errors (NMSE), for all the three maps on which it was tested. In particular, by choosing the optimal $\epsilon$ of the network for each task, lower errors are obtained in comparison with previous implementations in Ref. \cite{chen2020temporal}.
	
The lowest error in the tasks is obtained in correspondence of the $\epsilon$ values for which the MC is maximum.
	
	
The paper is organized as follows. In Section \ref{sect:method} we summarize the theory behind QRC and in particular quantum echo-state network while in Section \ref{sect:experiments} we describe the actual implementation as a circuit on a gate model quantum computer. The results are shown in Section \ref{sect:results} along with the detailed descrip\-tion of the tasks and the comparison between different settings for the reservoir with various values of the reset rate and numbers of qubits. The conclusions are drawn in the last Section \ref{sect:conclusion}.

\section{Method}\label{sect:method}
The key idea of QRC resides in using a $N-$ qubits quantum register as reservoir for an ESN. In such a way it is possible to exploit the high dimensionality of the Hilbert space associated to $N$ qubits, which effectively acts as a reservoir with a large number of nodes. Specifically, the nodes of the reservoir can be associated to the basis elements of the operator space, which for a $N$ qubits system amount to $4^N$ elements. 
We consider as basis elements the $N$-qubit Pauli operators defined as: 
\begin{equation}\label{pauliop}
P_i \equiv P_{i_{1} i_{2}\ldots i_{N}} = \bigotimes_{k=1}^{N} \sigma_{i_{k}}\quad,\qquad i_k\in\{00,01,10,11\}
\end{equation}
where $\{\sigma_{01},\sigma_{10},\sigma_{11}\}=\{X,Y,Z\}$ are the three Pauli matrices while $\sigma_{00}=I$, the $2\times2$ identity matrix, and $i$ is a natural number described by a binary string ranging from $000...0$ to $111...1$, build from the indexes of the Pauli matrices in the tensor product.

For a $N$ qubit reservoir, a single operator $P_i$ is the tensor product between $N$ matrices $\sigma\in \{\sigma_{01},\sigma_{10},\sigma_{11},\sigma_{00}\}$. There are $4^N$ possible total combina\-tions of such matrices, so that $i=1, 2, ...4^N$.  Thanks to the orthogonality property
\begin{equation}
    \mathrm{Tr}(P_iP_j) = 2^N\delta_{ij}
\end{equation}
a generic density matrix $\rho$ describing the quantum state of the system can be written as 
\begin{equation}
    \rho = \sum_{i=1}^{4^N} x_i\,P_i
\end{equation}
where the real numbers 
\begin{equation}\label{node}
    x_i=2^{-N}\mathrm{Tr}(P_i\rho)
\end{equation}
can be regarded as classical coordinates, or \emph{nodes}, of our qESN. Assuming $P_{4^N}=\bigotimes_{k=1}^{N} I_{k}$, one observes  that the last element of Eq. (\ref{node}) is constrained by normalization, i.e. $x_{4^N}=1$. Therefore, for a system of $N$ qubits there are $4^N - 1$ internal nodes of the reservoir.

We are interested in the learning of a temporal tasks where the input signal is an ordered array  $\{u(k)\}_{k=1}^{L}$ of $L$ real values with $u(k)\in[0,1]$ for every $k$. The input can be injected into the quantum reservoir with a pure states encoding, for example by initializing at each time step the first qubit of the reservoir in the state $|\psi_1\rangle = \sqrt{u(k)}|0\rangle + \sqrt{1-u(k)}|1\rangle$ \cite{PhysRevApplied.8.024030}. Instead, we follow the approach of Ref.\cite{chen2020temporal} so to encode the input as a classical mixture. Such an approach corresponds to initializing an ancilla qubit in the diagonally mixed state:
\begin{equation}\label{ancilla}
	\rho_a(u(k)) = u(k)|0\rangle\langle0| + (1-u(k))|1\rangle\langle1|
\end{equation} 
The ancilla qubit is then coupled with the $N-$qubits register of the reservoir in order to evolve the system with an input dependent map $T$ acting on the density matrix $\rho_R$ of the reservoir, initialized at $t=0$ in the pure state $\rho_R(0)=|0\rangle\langle0|^{\otimes N}$. In particular, at each time step $t=k$ the density matrix $\rho_R$ evolves as:
\begin{equation}\label{matrix_evo}
	 \rho_R(k) = T(u(k))\rho_R(k-1)
\end{equation} 
where $T(u(k))$ is a map which acts within the space of density matrices (i.e. $T:  \mathbb{C}^N\times\mathbb{C}^N\rightarrow  \mathbb{C}^N\times\mathbb{C}^N$) whose structure is described in Section \ref{sect:experiments}.
The time evolution of the reservoir serves the purpose of nonlinearly expanding the input signal, which is a sequence of scalar values, in a higher dimensional space, i.e. the reservoir degree. Indeed, by choosing a suitable map $T(u(k))$, the output signals extracted from the reservoir are nonlinear functions of the input values \cite{PhysRevApplied.8.024030}, \cite{fujii2021quantum}. In this way the network is able to emulate a nonlinear map by just adding a linear readout at its output.

After each injection of an input value and one--step evolution of the reservoir state, $N$ signals are extracted from the system and passed to the linear readout layer. The expectation values of the $Z$ operators i.e. $\{\langle Z_i\rangle\}_{i=1}^{N}$ provide the output signals. Such averages are computed as $\langle Z_i\rangle= \text{Tr}[Z_i\rho]$ and correspond to $N$ so--called \textit{true nodes} of the reservoir. 
The remaining $4^N-1-N$ so-called \textit{hidden nodes} are never measured. 
In Figure \ref{fig:Figure1} the structure of the qESN employed in the experiments is shown. In the next subsection the process of training the qESN and how the weights are updated is explained.

\afterpage{
\begin{figure*}
\centering
  \includegraphics[width=0.9\linewidth]{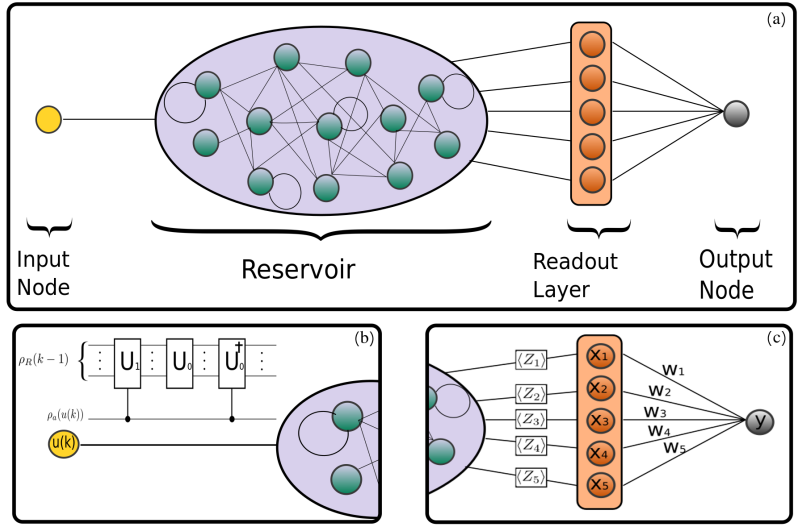}
  \caption{Structure of the qESN for a 5 qubits reservoir. Each node of the network (circles in the reservoir) does not correspond to a qubit, instead they are related to the basis operators $B_i$ by the Eq.\ref{node}. (a) Scheme of the qESN. At each time step an input value is injected into the $N=5$ qubits reservoir and one step of the time evolution of the system is performed. Right after, 5 signals are extracted from the reservoir and collected in the readout layer. Finally, the 5 signals are linearly combined to obtain the final output of the network.   (b) Injection of the input into the reservoir. At each time step an ancilla qubit is prepared in the mixed state $\rho_a(u(k)) = u(k)|0\rangle\langle0| + (1-u(k))|1\rangle\langle1|$. The ancilla is then coupled with the reservoir state $\rho_R(k-1)$ with controlled operators $U^\dagger_0$ and $U_1$ and the reservoir state evolves under $U_0$ or $U_1$ with a probability proportional to $u(k)$ and $(1-u(k))$ respectively, as required by Eq. (\ref{density_evo}). (c) Output extraction. After the one step evolution 5 signals are extracted from the reservoir by measuring $\langle Z\rangle$ on each qubit. The values of the 5 true nodes $x_i$ for $i=1,...,5$ are collected in the classical layer as rows of the matrix $\mathbf{X}$ as discussed in Section \ref{sect:training}. The final output is then obtained by a linear combination of the reservoir signals with the output weights. }
  \vspace{0.55cm}
  \label{fig:Figure1}
\end{figure*}

\begin{figure*}
\centering
\begin{subfigure}{.40\textwidth}
\vspace{4mm}
  \includegraphics[width=1.0\linewidth]{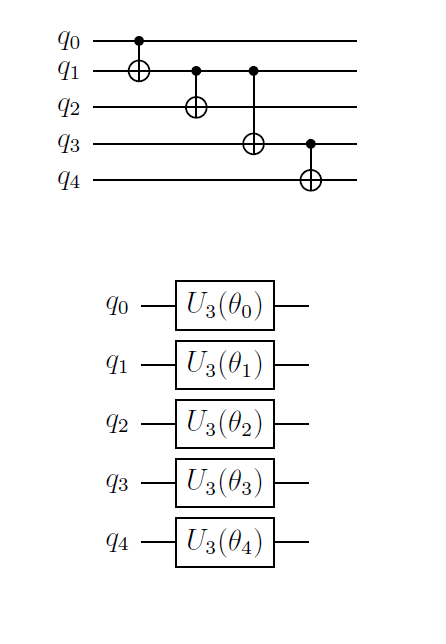}
\vspace{4mm}
\caption{Quantum circuits for $U_0$ and $U_1$ for the 5 qubits reservoir}
\end{subfigure}
\qquad\qquad
\begin{subfigure}{.40\textwidth}
  \includegraphics[width=1.0\linewidth]{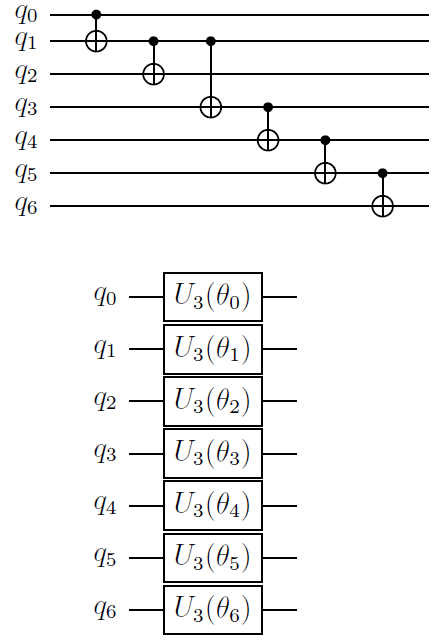}
\centering
\vspace{7mm}
\caption{Quantum circuits for $U_0$ and $U_1$ for the 7 qubits reservoir}
\end{subfigure}
\caption{
Circuit representation of the two unitary operator $U_0$ and $U_1$ used for the evolution of the 5 and 7 qubits reservoirs. The angles $\theta_0=(4.26,-1.14,0.198),\; \theta_1=(-1.84,3.54,-2.07),\; \theta_2=(5.34,0.186,2.96),\; \theta_3=(-3.31,4.03,-3.7),\;
\theta_4=(3.69,-3.84,-3.92)$ for $U_1$ are taken from \cite{chen2020temporal}, while the angles $\theta_5=(-2.21, 3.04, 2.5)$ and $\theta_6=(1.69, -2.34, 2.51)$ are chosen randomly.}
\label{fig:Figure2}
\end{figure*}

}

\subsection{Training}\label{sect:training}
In the following, the goal of the temporal task is to learn a non linear map $f:\mathbb{R}\rightarrow\mathbb{R}$ which, given the input values $\{u(k)\}_{k=1}^{L}$, returns the output signal $\{\bar{y}(k)\}_{k=1}^{L}$ with $\bar{y}(k)\in \mathbb{R}$. Since in general the output at a given time depends also on the history of the input signal at previous times, that is $\bar{y}(k)=f(\{u(i)\}_{i=0}^{k})$, some kind of memory mechanism is required on the network. Indeed the quantum evolution in Eq. (\ref{matrix_evo}) allows the system to retain information about past inputs, so it provides the ability to correctly emulate the given map $f$. In the following, three different maps are considered, all of them with fading memory, i.e. each output depends increasingly less on previous inputs as time proceeds. Further details on them are given in Section \ref{sect:tasks}.

The basic idea of reservoir computing is to make a clear--cut  distinction between the reservoir and the linear readout and train them differently. Specifically, once all hyperparameters are properly set, the only weights which are adjusted during the training in an ESN are those of the linear readout, through simple linear regression.

Here, for every $k^{th}$ time step, $N$ signals are extracted from the reservoir: let's denote by $x_{ki}$ the value on the $i-$th true node at the time $t=k$, with $i=1,2,...,N$ and $k=1,...,L$, where $L$ is the total temporal length of the input signal. The $x_{ki}$ form the  $L\times N$ design matrix $\mathbf{X}$.
As target values for the training and testing we use the $\{\bar{y}(k)\}_{k=1}^{L}$ signal obtained from the maps for the corresponding input $\{u(k)\}_{k=1}^{L}$. These target values can be collected in a $L\times 1$ array $\mathbf{Y}$. 
So, at the end, linear regression is performed over the matrix equation $\mathbf{Y}=\mathbf{X}\mathbf{W}$ calculating the $N$ trained weights as:
\begin{equation}\label{weight}
\mathbf{W}^\mathrm{trained}=\mathbf{X}^{\dagger}\mathbf{Y}
\end{equation}
where $\mathbf{X}^{\dagger} = (\mathbf{X}^{T}\mathbf{X})^{-1}\mathbf{X}^T$ is the Moore-Penrose pseudo-inverse of $\mathbf{X}$. Once the weights are trained, the prediction values for $k>L$  are computed by:
\begin{equation}
    \mathbf{Y}^\mathrm{predicted}=
    \mathbf{\tilde{X}}\mathbf{W}
    ^\mathrm{trained}
\end{equation} 
where $\mathbf{\tilde{X}}$ collects the output values when $k=L+1,L+2\ldots$. We recall that these output signals are read from a  reservoir whose internal dynamics is not subject to any training procedure, as appropriate to an Echo State Network.
\section{Experiments}\label{sect:experiments}
In this Section we discuss the details of the actual implementation of the qESN. For the quantum part of the network, i.e. the reservoir, we employed both a 5 and a 7 qubit system, the former in order to compare the results with those in Ref.\cite{chen2020temporal}. We implemented it as a circuit on both the IBM Qiskit simulator and quantum hardware \texttt{ibmq\_casablanca}. 
For the output layer, instead, a classical linear readout with $N$ input and one output nodes was used, which we implemented by linear regression using the Python library \texttt{sklearn}.

As discussed in Section \ref{sect:method} at every time step the system evolves under a map $T(u(k))$ which depends on the input value $u(k)$ at that time step $t=k$. For the sake of consistency with previous literature, the stochastic evolution map $T$ proposed in Ref. \cite{chen2020temporal}, acting on the density matrix of the reservoir, is considered, by:
\begin{equation}\label{density_evo}
\begin{split}
\rho(k) &= T(u(k))\rho(k-1) \\
&=\begin{cases}
T_0\rho(k-1) \;\; \text{w.p. $(1-\epsilon)u(k)$}\\
T_1\rho(k-1) \;\; \text{w.p. $(1-\epsilon)(1-u(k))$}\\
\sigma \qquad \qquad \;\;\text{w.p. $\epsilon$}
\end{cases}
\end{split}
\end{equation} 
where "w.p." is a short hand for "with probability". Here $0\le u(k)\le1$ is the input at the time $t=k$ while $T_0$ and $T_1$ are two arbitrary completely positive trace-preserving (CPTP) maps. Hence they can be implemented with two unitary operators $U_0$ and $U_1$ according to the usual scheme for time evolution of density matrices in Quantum Mechanics:
\begin{equation}
T_j\rho=U_j\,\rho\,U_{j}^{\dagger}\;,\quad j=0,1.
\end{equation}
In the third alternative of Eq.~\eqref{density_evo}, the quantum state is reset to some fixed density operator $\sigma$ with probability $\epsilon$ where  $0\le \epsilon\le 1$. The hyperparame\-ter $\epsilon$, which can be called \emph{reset rate}, allows the qESN to have a fading memory. If $\epsilon=1$ there is no evolution and no dependence whatsoever on the inputs $\{u(k)\}_{k=1}^{L}$. At the opposite end, when $\epsilon=0$, the network undergoes a mixture of unitary evolution steps. The net evolution is not unitary, but it might retain too much information on the initial state, hindering the driving capability of the inputs.

A particular history of the stochastic process in Eq. (\ref{density_evo}) corresponds to a quantum circuit where we encode the input on an ancilla qubit in a mixed state, as in Eq. (\ref{ancilla}). Then we couple the ancilla with the reservoir through controlled unitary operators $U_j$ as shown in Figure \ref{fig:Figure1}(b).
To implement the evolution map (\ref{density_evo}) in Qiskit, which does not allow for mixed-state qubit preparation, we employ an ensemble description \cite{chen2020temporal}. Specifically we reproduce the mixed state encoding of the input by preparing an ensemble of quantum circuits all in the same initial pure state $|\psi(0)\rangle = |0\rangle^{\otimes N}$ (with $N=5$ or $N=7$)  and evolving them with operators $U_0$ or $U_1$ with the corresponding probability law. 
In particular we prepare an ensemble of $N_c$ circuits in the all spin up state and at every time step $t=k$ we evolve each of them by applying either one of the two unitary operators $U_0$ or $U_1$, or we reset the state $|\psi(k)\rangle$ to $|\psi(0)\rangle$:
\begin{equation}\label{evolution}
|\psi(k)\rangle=\begin{cases}
U_0|\psi(k-1)\rangle \;\;\;\; \text{w.p. $(1-\epsilon)u(k)$}\\
U_1|\psi(k-1)\rangle \;\;\;\; \text{w.p. $(1-\epsilon)(1-u(k))$}\\
|\psi(0)\rangle \qquad \qquad \text{w.p. $\epsilon$}
\end{cases}
\end{equation}
The two unitary operators $U_0$ and $U_1$ can be chosen arbitrarily, provided they do not produce trivial dynamics, as would be for example if $U_0 \propto U_1$ and do generate a nonlinear dependence on $\{u(k)\}_{k=1}^{L}$. To ensure that the next results can be compared with prior art, the implementation follows the same choice already discussed in Ref. \citep{chen2020temporal}. The circuits for the two unitary operators (shown in Figure \ref{fig:Figure2}) consist of $U_0$ as a highly non trivial operator composed only by CNOT gates, while $U_1$ takes into account 3D spin rotations on each qubit, respectively.

Summarizing, $N_c$ circuits of $N-$qubits (with $N=5$ or $N=7$) each are created and initialized, by default, at the state $|0\rangle^{\otimes N}$. At each time step $t=k$, for each one of such circuits, the sequence of gate $U_0$ or $U_1$ is applied accordingly to the input dependent probabilities as from Eq. (\ref{evolution}). With a probability $\epsilon$, instead, each circuit is reset in the initial state $|0\rangle^{\otimes N}$. For each one of the $N_c$ circuits, after every step evolution, the measurement of the $Z_j$ operators is done. Such process is repeated for all the $L$ time steps and the results are averaged over all the $N_c$ circuits. We finally obtain the design $L\times N$ matrix $\mathbf{X}$ which collects all output histories from the reservoir. Such matrix is then used as argument of the linear readout to finally obtain the output signal of the ESN and compute the trained weights $\mathbf{W}^{trained}$ with linear regression as in Eq. (\ref{weight}). In the next section, the details of the tasks used to test our qESN are described.

\subsection{Description of the tasks}\label{sect:tasks}
In this Section the three tasks over which the qESN has been tested are explained. The three tasks are taken from the ones employed in Ref. \cite{chen2020temporal}, in such way it was possible to compare the results. All of them consist in learning a nonlinear map, that means to induce the system to emulate a selected map predicting the correct outputs for input values not seen before.

For all the three tasks the input is a length $L$ array of random values extracted in $[0,1]$ where the $k-th$ entry is  $u(k)$ of from the previous Sections.
In tasks I and II we employ two maps that process the input by updating linearly an internal highly dimensional state $\mathbf{z}$ at each time step $k$; then the map's output value is given by a polynomial readout $h$:
\begin{equation}\label{task_map}
\mathbf{z}(k)=A\mathbf{z}(k-1)+u(k)\mathbf{c}\;,\quad\bar{y}(k)=h(\mathbf{z}(k))
\end{equation}
Here $\mathbf{z},\mathbf{c}\in R^{n}$ with $n=2000$ and $\mathbf{c}$ an arbitrary fixed array, while $h$ is a polynomial function, in our case taken to be of second degree, which returns a single scalar value. Specifically, $\mathbf{c}$ and all the parameters of the $h$ function were selected randomly in $[-1,1]$, while the initial condition $\mathbf{z}(0)$ was chosen to be identically null. 
The two maps differ for the structure of the matrix $A\in R^{n\times n}$: while in the first case we used a dense matrix with a maximum singular value of $\sigma(A)_{max}=0.5$, in the second case a 95\% sparse matrix with $\sigma(A)_{max}=0.99$ was employed.

The third task instead relates to the simplified dynamics of an aircraft moving in the horizontal plane. In particular, if the velocity is constant then it can be shown \cite{ni1996new} that such dynamics is defined by the couple of differential equation:
\begin{align}
 \dot{x}_1 &= x_2 -0.1(5x_1-4x_1^3+x_1^5)\text{cos}x_1-0.5\Tilde{u}\;\text{cos}x_1 \\
 \dot{x}_2 &=-65x_1 +50x_1^3 - 15x_1^5-x_2-100\;\Tilde{u}
 \end{align} 
Where $\Tilde{u}$ represents the rudder deflection at a given time which, as pointed out in Ref. \cite{ni1996new}, should be in the range $[-0.5,0.5]$. So at each time $t=k$ the relation between $\Tilde{u}$ and the input values $u(k)$ is $\Tilde{u}=0.5u(k)-1$. The two variables $x_1$ and $x_2$, instead, concern the slideslip and yaw angle respectively. The output is then taken to be $\bar{y}(k)=x_1(k)$ and the target values are obtained using the \texttt{ode45} solver in MATLAB with a sampling time of $t=1$.

For all three tasks, the target signal is the output array with entries $\{\bar{y}(k)\}_{k=1}^{L}$, given the input $\{u(k)\}_{k=1}^{L}$. Specifically, we set $L=60$ in the experiments on the Qiskit simulator while $L=30$ on the quantum hardware for time taking reasons, as we will explain in the next Section.

\begin{figure*}[t!]
  \includegraphics[width=\linewidth]{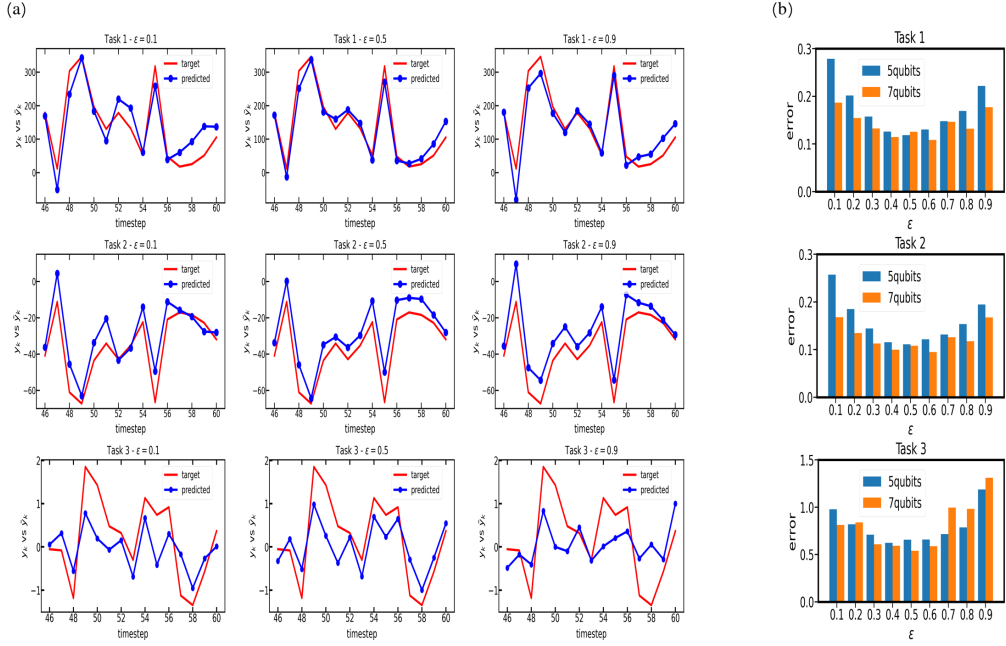}
  \caption{Results on the simulator. (a) Comparison between the output $y_k$ predicted from the 5 qubits qESN and the target values $\bar{y}_k$ obtained from the map for the 15 test values of a single input signal. For each task we show the results for $\epsilon=[0.1, 0.5, 0.9]$. The qESN performs definitely better in the first two tasks while in the third task the output signal follows the expected one only for $\epsilon=0.5$. (b) Histograms of the average errors obtained in the different tasks using a 5 and 7 qubits reservoirs. For tasks I and II the lowest error with the 5 qubits reservoir is obtained for $\epsilon = 0.5$ while $\epsilon= 0.4$ for Task III.}
  \vspace{2cm}
  \label{fig:Figure3}
\end{figure*}

\section{Results}\label{sect:results}
In this Section we show the results obtained for the three tasks both on the simulator and on the quantum hardware. Since quantum measurements perturb the state, after the extraction of the reservoir signal the information stored in the qubits is lost and the qubit is no longer usable. To circumvent this problem, after the measurement at $t=k$ every circuit is reinitialized to $|0\rangle^{\otimes N}$ and left to evolve unperturbed until $t=k+1$, where a new measurement is performed. Therefore, to process an input signal of length $L$, one needs to reinitialize the circuit $L-1$ times and for each time step to increase by 1 the number of applications of the evolution in Eq. (\ref{evolution}), requiring a total of $1+2+...+L=L(L+1)/2$ applications of the gates in Eq. (\ref{evolution}). Furthermore, every circuit at every time step is run for a total number $S$ of shots. Since, as we said, for the Monte Carlo sampling we employ $N_c$ circuit copies, the total number of circuits runs and application of Eq. (\ref{evolution}) are $N_cSL$ and $N_cS(L+1)L/2$   respectively. In order to reduce the  computational time, we decided to take in consideration shorter input signal of $L=30$ and fewer circuit copies $N_c=300$ for the experiments on the quantum hardware compared to the $L=60$ and $N_c=1024$ employed on the simulator.
All the parameters of the experiments are summarized in the Table \ref{tab:param}.

\begin{table}[h!]
	\centering
	\captionsetup{font=sc,skip=6pt, justification = centering,labelfont=large,labelfont=bf, labelsep = newline, singlelinecheck=false }
	\caption{parameters used in the experiments on the simulator and on the quantum computer}
	\setlength{\tabcolsep}{15pt}
	\small
	\begin{tabular}{p{4cm}p{1.0cm}p{1.3cm}p{1.4cm}}
	
		\toprule[.4mm]
		Parameter &Symbol& Simulator & Quantum hardware\\
		\midrule[.2mm]
		Circuit copies& $N_c$ & 1024	    &  300\\
		Number of shots& $S$  & 4000     &  4000\\
		Input length& $L$     & 60       &   30\\
		\bottomrule[.4mm]
		
	\end{tabular}
	\label{tab:param}
\end{table}

\begin{figure*}[t!]
  \includegraphics[width=\linewidth]{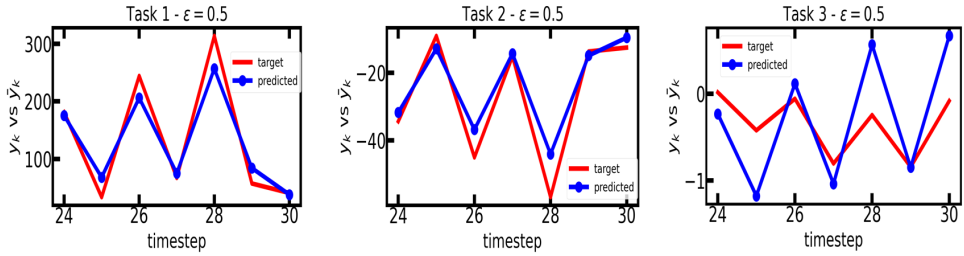}
  \caption{Results on the IBM quantum computer \texttt{ibmq\_casablanca} for the 5 qubits reservoir. Comparison between the output $y_k$ predicted from the qESN and the target values $\bar{y}_k$ obtained from the map for the 7 test values of a single input signal.}
  \vspace{2cm}
  \label{fig:Figure4}
\end{figure*}

\subsection{Results on the simulator}\label{sect:res_simu}
For the experiments on the simulator we extracted $L=60$ random values in $[0,1]$ which formed the entries of the input array. In particular, after discarding the first 10 values, the training was performed over the next 35 inputs, i.e. until the time step $L_T=45$. Such values were used to adjust the readout weights by linear regression. The last 15 values, which correspond to the time steps $k= L_T+1, ..., L$, were employed for the testing where we compared the output from the qESN to the $\{\bar{y}(k)\}_{k=46}^{60}$ expected outputs obtained from the map of the given task. To asses the goodness of the performance we measured the discrepancy between the expected $\bar{y}(k)$ and obtained $y(k)$ output by computing a normalized mean square error (NMSE):
\begin{equation}
\mathit{E}=\sum_{l=L_T+1}^{L}|y(l)-\bar{y}(l)|^2/\Delta_y^2
\end{equation}
where $\Delta_y^2=\sum_{l=L_T+1}^{L}(y(l)-\mu)^2$ with $\mu=1/(L-L_T)\sum_{l=L_T+1}^{L}y_l$.\\
We performed each task for different values of $\epsilon\in\{0.1,0.2,0.3,0.4,0.5,0.6,\\0.7,0.8,0.9\}$, always keeping the same input signal, in order to investigate which one performs better and then use it on the quantum hardware. The experiments were repeat 15 times with different inputs and the final errors of the testing were averaged. In Figure \ref{fig:Figure3}(a) we show the results obtained in a single experiment for each of the three tasks while in Figure \ref{fig:Figure3}(b) the averaged errors evaluated over the 15 different inputs are plotted. The lowest errors are obtained with $\epsilon = 0.5$ for Task I and II while $\epsilon=0.4$ for Task III. Figure \ref{fig:Figure3}(b) shows the average errors for a reservoir made of 7 qubits. Its evolution operators $U_0$ and $U_1$ are based on the same structure of those used with the 5 qubits case and adapted them to a 7 qubits register, as shown in Figure \ref{fig:Figure2}. 
Except for Task III, the 7 qubits reservoir returns a lower error for every $\epsilon$ values. For all the three Tasks the optimal $\epsilon$ of the 7 qubits reservoir slightly differs from the 5 qubits case ($\epsilon=0.6 $ vs $\epsilon=0.5$ for Tasks I and II, $\epsilon=0.5$ vs $\epsilon=0.4$ for Task III). The following Section deals with quantum hardware with a reservoir of 5 qubits and therefore we selected the optimal $\epsilon$ value accordingly.

\subsection{Results on quantum hardware}
For the experiments on quantum hardware we employed the IBM quantum computer \texttt{ibmq\_casablanca}. As shown in Table \ref{tab:param} we used an input of $L=30$ values, in this case the training was performed over the first $L_T=23$ entries of the input array (the effective training was performed over 19 values as the first 4 inputs were discarded), while the last 7 were used for the testing. We run our experiment using an $\epsilon$ value of $\epsilon=0.5$ which is the best for Task I and II and still gives low error for Task III, as seen from  Figure \ref{fig:Figure3}(b).  The results are shown in Figure \ref{fig:Figure4}, while the performances of the network in each task are summarized in the Table \ref{tab:res_real}.
With respect to prior art, which is represented by the implementation of the same operators on the 5 qubits Ourense quantum computer of IBM at $\epsilon=0.1$ \cite{chen2020temporal}, we improve the MSE from 0.24 to 0.09 for Task I, from 0.68 to 0.12 for Task II, respectively, while the results on the Task III are comparable (MSE is 2.56 instead of 2.3), but one should keep in mind that we used less than one third of circuit copies. 
\begin{table}[h!]
	\centering
	\captionsetup{font=sc,skip=6pt, labelsep = newline, justification = centering,labelfont=large,labelfont=bf,singlelinecheck=false}
	\caption{results obtained in the three tasks on the quantum hardware}
	\setlength{\tabcolsep}{20pt}
	
	\begin{tabular}{llll}
		
		\toprule[.4mm]
		Task           &$\epsilon$ & MSE [This work] & MSE  \cite{chen2020temporal}\\
		\midrule[.2mm]
		I & 0.5	    &  0.09 &0.24\\
		II  & 0.5     &  0.12&0.68\\
		III     & 0.5      &   2.56&2.3\\
		\bottomrule[.4mm]
		
	\end{tabular}
	\label{tab:res_real}
\end{table}

\begin{figure*}[t!]
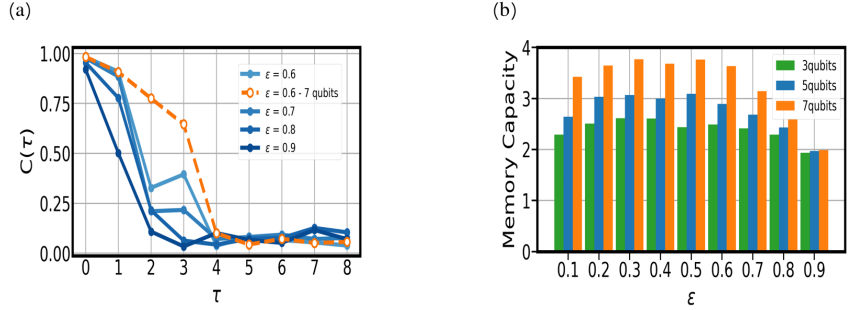

\centering
 	\includegraphics[width=0.4\linewidth]{Figure5a.pdf}
 	\qquad\includegraphics[width=0.4\linewidth]{Figure5b.pdf}
 	\caption{
 	Investigation of the memory capacity of the network. In (a) we plot the correlation $C(\tau)$ between the predicted output and the target $\{\bar{y}(k)\}_{k=45}^{60}=\{u(k-\tau)\}_{k=45}^{60}$, where $u(k)$ is the input value at the time $t=k$. We plot the correlation for different values of $\tau$ and compared the results for 4 different $\epsilon$ in the evolution of the reservoir for the 5 qubits reservoir (in blue). The correlation decreases, as expected, as $\tau$ increases. Furthermore, at higher $\epsilon$ the $C(\tau)$ decreases more rapidly. A slower decrease of $C(\tau)$ can be seen for the 7 qubits reservoir.
 	The histogram in (b) represents the MC at different $\epsilon$ for the 5 qubits reservoir (in blue). It is almost constant from $\epsilon = 0.2$ to $\epsilon=0.5$, then it starts to decrease. The comparison with the 3 and 7 qubits reservoirs shows an increase of the memory capacity with the number of qubits.}
 	\label{fig:Figure5}
 \end{figure*}
 
\subsection{Evaluation of the memory of the qESN} \label{sect:memory}
Finally, we present the dependence of the memory of the qESN on the reset rate $\epsilon$. The memory of the system is quantitatively described by the memory capacity (MC) metric \cite{PhysRevApplied.8.024030}. Such quantity determines the ability of the qESN to reproduce at a time $t$ inputs received at the earlier time $t-\tau$. As $\tau$ increases the reservoir should gradually loose information about its past, so the accuracy of the predicted output should decrease. We run our experiment on the Qiskit simulator. In particular we prepared an input array $u=\{u(k)\}_{k=15}^{60}$ with random values $u(k)\in[0,1]$ and we injected into the qESN the first $45$ values as training set, picking as target array the same sequence shifted by a delay $\tau$, i.e. $u_\tau=\{u(k-\tau)\}_{k=15}^{60}$. After discarding the first 14 output values, we trained the network by comparing output and target for the times $k=15,...,45$, while the last 15 values were used for the testing as we did in the three tasks in Section \ref{sect:res_simu}.
We then calculated the correlation between the output signal of the network $y=\{y(k)\}_{k=45}^{60}$ and the target $t_\tau$ for a given $\tau$:
\begin{equation}
    C(\tau)=\frac{\text{COV}^2(y,u_\tau)}{\sigma^2(y)\sigma^2(u_\tau)}
\end{equation}
where COV$(x,y)$ and $\sigma(x)$ represent the covariance between $x$ and $y$ and the standard deviation of $x$, respectively.  
The memory capacity $M_c$ is then computed by summing the values of $C(\tau)$ for different $\tau$:
\begin{equation}
    M_c = \sum_{\tau=0}^{\tau=\tau_\text{max}}C(\tau)
\end{equation}
where in our case $\tau_\text{max}=8$.
The results for the memory capacity tests are discussed in the next Section. In Figure \ref{fig:Figure5}(b) the histograms representing the memory capacity at different $\epsilon$'s were displayed for reservoirs of 3, 5 and 7 qubits. In Figure \ref{fig:Figure5}(a), instead, the behavior of $C(\tau)$ as $\tau$ varies is shown for various $\epsilon$'s.

To test the memory capacity, we employed three different reservoirs of 3, 5 and 7 qubits, respectively. For each one, the performance for different values of the reset rate $\epsilon$ were evaluated.

In the case of the 3 qubits reservoir we adopted for the evolution operator $U_0$ a circuit with one CNOT between the first-second and second-third quibits while for $U_1$ we took the same of Figure \ref{fig:Figure2} considering only the first three qubits. In such way we were able to compare the memory of the three different reservoirs. The results in Figure \ref{fig:Figure5} suggest that for the optimal $\epsilon=0.5$ the increase in the value of the MC (2.43, 3.09, 3.76, for $N=3,5,7$ respectively) with the number of qubits is approximately linear, confirming previous results \cite{PhysRevApplied.8.024030} with a different implementation of a quantum reservoir. Such result is expected as increasing the number of qubits will increase linearly the number of observables we measure in the output extraction process, i.e. the $\langle Z \rangle$ expectation values measured on each qubit.

\section{Conclusions}\label{sect:conclusion}
We implemented a quantum Echo-State Network on a gate model quantum computer and tested it over three different tasks consisting to predict outputs of nonlinear maps, to maximize its performances. 
We investigated the memory of the reservoir by evaluating its memory capacity for different values of the reset rate  $\epsilon$. The memory capacity of the reservoir 
remains stable at its highest values for the central values of $\epsilon\in [0,1]$, while it starts to decrease for $\epsilon>0.5$, as expected since the reservoir resets its internal state more often.
After choosing the best $\epsilon$ parameter for the reservoir evolution based on the simulator at 0.5, the results obtained on the quantum hardware show low NMSE. In particular, by comparing the MSE values with those reported in literature at $\epsilon=0.1$, we observe an improvement of 62\% and 82\% in Task I and II respectively. For the task involving the pair of differential equations, a comparable error is found by using less than one third of circuit copies.For completeness, it should be noted that the results in Ref. \cite{chen2020temporal} are obtained on a different machine (Ourense) than the one employed in this work (Casablanca) as Ourense is no longer available on the IBM system. The results obtained might, of course, be affected by the different nature of noise in the two machines. However, the decrease of MSE for central values of $\epsilon\in [0,1]$ is not influenced by the nature of the machines and, so, by choosing an optimal $\epsilon$ better results than Ref. \cite{chen2020temporal} are in any case expected.
We confirm that also for the optimal $\epsilon=0.5$ the memory capacity grows approximately linearly with the number of qubits. Such result is consistent with that obtained in Ref. \cite{PhysRevApplied.8.024030} where a different reservoir evolution was considered and the implementation was done by a simulated NMR machine.
There is no direct proportionality between the memory capacity of the network and its errors in the tasks. However, the best performances are found for the $\epsilon$ values for which the memory capacity is highest while the errors in the tasks start to increase as the memory capacity decreases at the largest values of $\epsilon$.
Our investigation shows that, in order to maximize the performances of the qESN, it is convenient to tune the hyperparameters of the quantum reservoir such that the network develops a high memory capacity. 

We also notice that the optimal $\epsilon$ found in our work should not in principle depend on the task in consideration. In fact the dynamics of the reservoir can be regarded as that of many chaotic oscillators with frequencies which spread over several orders of magnitude. In that sense, the optimal setting of the reservoir should not be overly influenced by the time dependence that the function to be emulated sets over input data. On the other hand, we do expect the weights of the readout layer do strongly depend on such time scale.


\begin{thebibliography}{10}
\expandafter\ifx\csname url\endcsname\relax
  \def\url#1{\texttt{#1}}\fi
\expandafter\ifx\csname urlprefix\endcsname\relax\def\urlprefix{URL }\fi
\expandafter\ifx\csname href\endcsname\relax
  \def\href#1#2{#2} \def\path#1{#1}\fi

\bibitem{rotta2017quantum}
D.~Rotta, F.~Sebastiano, E.~Charbon, E.~Prati, Quantum information density
  scaling and qubit operation time constraints of {CMOS} silicon-based quantum
  computer architectures, npj Quantum Information 3~(1) (2017) 1--14.

\bibitem{ferraro2020all}
E.~Ferraro, E.~Prati, Is all-electrical silicon quantum computing feasible in
  the long term?, Physics Letters A 384~(17) (2020) 126352.

\bibitem{rocutto2021quantum}
L.~Rocutto, C.~Destri, E.~Prati, Quantum semantic learning by reverse annealing
  of an adiabatic quantum computer, Advanced Quantum Technologies 4~(2) (2021)
  2000133.

\bibitem{maronese2021continuous}
M.~Maronese, E.~Prati, A continuous {Rosenblatt} quantum perceptron,
  International Journal of Quantum Information (2021) 2140002.

\bibitem{agliardi2022optimal}
G.~Agliardi, E.~Prati, \href{https://www.mdpi.com/2624-960X/4/1/6}{Optimal tuning of quantum generative adversarial networks for multivariate
  distribution loading}, Quantum Reports 4~(1) (2022) 75--105.
\newblock \href {https://doi.org/10.3390/quantum4010006}
  {\path{doi:10.3390/quantum4010006}}.
\newline\urlprefix\url{https://www.mdpi.com/2624-960X/4/1/6}



\bibitem{PhysRevApplied.8.024030}
K.~Fujii, K.~Nakajima, Harnessing disordered-ensemble quantum dynamics for machine learning, Phys. Rev. A 8 (2017) 024030.

\bibitem{Soriano}
J.~Nokkala, R.~Mart{\'\i}nez-Pe{\~n}a, R.~Zambrini, M. ~C. Soriano, High-Performance Reservoir Computing With Fluctuations in Linear Networks,
IEEE Transactions on Neural Networks and Learning Systems 33 (2022) 032315.
\newblock {\path{doi:10.1109/TNNLS.2021.3105695}}.

\bibitem{agliardi2022quantum}
G. ~Agliardi, M. ~Grossi, M.~Pellen, E.~Prati, Quantum integration of elementary particle processes, Physics Letters B (2022) 137228.


\bibitem{maass2004computational}
W.~Maass, M.~Henry, On the computational power of circuits of spiking neurons, Journal of computer and system sciences 69 (2004).


\bibitem{nakajima2019boosting}
K.~Nakajima, K.~Fujii, M.~Negoro, K.~Mitarai, M.~Kitagawa, Boosting computational power through spatial multiplexing in quantum reservoir computing, Physical Review Applied 11 (2019) 034021.



\bibitem{maronese2022quantum}
M.~Maronese, C.~Destri, E.~Prati, Quantum activation functions for quantum neural networks, Quantum Inf Process 21 (2022).


\bibitem{lazzarin2022multi}
M.~Lazzarin, D.~Galli, E.~Prati, Multi-class quantum classifiers with tensor network circuits for quantum phase recognition, Physics Letters A (2022) 128056.



\bibitem{dasgupta2020designing}
S.~Dasgupta, K.~Hamilton, A.~Banerjee, Designing a NISQ reservoir with maximal memory capacity for volatility forecasting, arXiv preprint arXiv:2004.08240 (2020).


\bibitem{fujii2021quantum}
K.~Fujii, K.~Nakajima,  Quantum reservoir computing: a reservoir approach toward quantum machine learning on near-term quantum devices, Springer (2021).




\bibitem{chen2020temporal}
J.~Chen, H.~Nurdin, N.~Yamamoto, Temporal information processing on noisy quantum computers, Physical Review Applied 14 (2020) 024065.


\bibitem{lukovsevivcius2009reservoir}
M.~Luko{\v{s}}evi{\v{c}}ius, H.~Jaeger,  Reservoir computing approaches to recurrent neural network training, Computer Science Review 3 (2009).


\bibitem{ni1996new}
Xianfeng.~Ni, M.~Verhaegen, A.~J.Krijgsman, H.~B.Verbruggen, A new method for identification and control of nonlinear dynamic systems, Engineering Applications of Artificial Intelligence 9 (1996).


\bibitem{tran2021learning}
Q.~H. Tran, K.~Nakajima, Learning temporal quantum tomography, Physical review letters 127 (2021) 260401.


\bibitem{jaeger2004harnessing}
H.~Jaeger, H.~Haas, Harnessing nonlinearity: Predicting chaotic systems and saving energy in wireless communication, science 304 (2004) 5667.


\bibitem{jaeger2002tutorial}
H.~Jaeger,  Tutorial on training recurrent neural networks, covering BPPT, RTRL, EKF and the "echo state network" approach, GMD-Forschungszentrum Informationstechnik Bonn 5 (2002).


\bibitem{jaeger2001echo}
H.~Jaeger, The “echo state” approach to analysing and training recurrent neural networks-with an erratum note, Bonn, Germany: German National Research Center for Information Technology GMD Technical Report 148 (2001).


\bibitem{nielsen2010quantum}
M.~Nielsen, I.~L. Chuang, Quantum Computation and Quantum Information, Cambridge University Press (2010).


\bibitem{barends2014superconducting}
R.~Barends, J.~Kelly, A.~Megrant, A.~Veitia, D.~Sank, E.~Jeffrey, T.~C. White, J.~Mutus, A.~G. Fowler, B.~Campbell, Superconducting quantum circuits at the surface code threshold for fault tolerance, Nature 508 (2014) 7497.

\bibitem{preskill2018quantum}
J.~Preskill, Quantum computing in the {NISQ} era and beyond, {Quantum} 2 (2018)
  79.
\newblock \href {https://doi.org/10.22331/q-2018-08-06-79}
  {\path{doi:10.22331/q-2018-08-06-79}}.


\bibitem{boixo2018characterizing}
S.~Boixo, S.~Isakov, V.~Smelyanskiy, R.~Babbush, N.~Ding et al, Characterizing quantum supremacy in near-term devices, Nature Physics 14 (2018) 595-600.






\end{thebibliography}

\end{document}